\documentclass[12pt]{article}
\usepackage{psfig}
\usepackage{graphicx,floatflt,amssymb,epsf}
\usepackage{graphics}
\global\parskip 6pt

\normalsize

\let\oldtheequation=\theequation
\def\doteqs#1{\setcounter{equation}{0}
            \def\theequation{{#1}.\oldtheequation}}

\newcounter{sxn}
\def\sx#1{\addtocounter{sxn}{1} \bigskip\medskip \goodbreak
\noindent{\large\bf 
{\thesxn.~~#1}} \nobreak \medskip}
\def\sxn#1{\sx{#1} \doteqs{\thesxn}}

\newcounter{axn}

\def\br{\begin{thebibliography}{abc}}
\def\er{\end{thebibliography}}

\def\be{\begin{equation}}

\def\ee{\end{equation}}

\def\bea{\begin{eqnarray}}

\def\eea{\end{eqnarray}}

\begin{document}
\begin{flushright}
\hfill{SINP-TNP/06-09}\\
\end{flushright}
\vspace*{1cm}
\thispagestyle{empty}
\centerline{\large\bf Quasinormal modes for tensor and vector type perturbation of}
\centerline{\large\bf Gauss Bonnet black hole using third order WKB approach }
\bigskip
\begin{center}
Sayan K. Chakrabarti\footnote{Email: sayan.chakrabarti@saha.ac.in}\\
\vspace*{.5cm}
{\em Theory Division\\
Saha Institute of Nuclear Physics\\
1/AF Bidhannagar, Calcutta - 700064, India}\\
\vspace*{.5cm}
\end{center}
\vskip.5cm

\begin{abstract}
  
  We obtain the quasinormal modes for tensor perturbations of
  Gauss-Bonnet (GB) black holes in $d=5, 7, 8$ dimensions and vector
  perturbations in $d = 5, 6, 7$ and $8$ dimensions using third order
  WKB formalism. The tensor perturbation for black holes in $d=6$ is
  not considered because of the fact that it is unstable to tensor
  mode perturbations. In the case of uncharged GB black hole, for both
  tensor and vector perturbations, the real part of the QN frequency
  increases as the Gauss-Bonnet coupling ($\alpha'$) increases. The
  imaginary part first decreases upto a certain value of $\alpha'$ and
  then increases with $\alpha'$ for both tensor and vector
  perturbations. For larger values of $\alpha'$, the QN frequencies
  for vector perturbation differs slightly from the QN frequencies for
  tensorial one. It has also been shown that as $\alpha' \to 0$, the
  quasinormal mode frequency for tensor and vector perturbation of the
  Schwarzschild black hole can be obtained. We have also calculated
  the quasinormal spectrum of the charged GB black hole for tensor
  perturbations. Here we have found that the real oscillation
  frequency increases, while the imaginary part of the frequency falls
  with the increase of the charge. We also show that the quasinormal
  frequencies for scalar field perturbations and the tensor
  gravitational perturbations do not match as was claimed in the
  literature. The difference in the result increases if we increase
  the GB coupling.
\end{abstract}
\vspace*{.3cm}
\begin{center}
\end{center}
\vspace*{.3cm}
\newpage

\sxn{Introduction}

Quasinormal Modes (QNMs) are of great relevance in discussing the
perturbations of black holes \cite{kk,noll,rg,zerilli,vish,vish3}.
For a large class of black holes, the equations governing the
perturbations can be cast into the Schr\"{o}dinger like wave equation.
For asymptotically flat space-times, the QNMs are solutions of the
corresponding wave equation with complex frequencies which are purely
ingoing at the horizon and purely outgoing at spatial infinity
\cite{kk, noll}. QNMs associated with metric perturbations have been found
to be a useful probe of the underlying space-time geometry
\cite{rg,vish}.  QN frequencies carry unique information about black
hole parameters and are expected to be detected in future
gravitational wave detectors \cite{sterio}.  In addition, it has also
been shown that quasinormal modes in Anti-de Sitter space-time appear
naturally in the description of the corresponding dual conformal field
theories living on the boundary \cite{danny, danny1}. In spite of their
classical origin, it has also been suggested that QNMs might provide a
glimpse into the quantum nature of the black hole \cite{hod, dreyer,
  motl, sdas, sdas1, natario}.

QNMs of black holes have been traditionally studied for the
backgrounds which arise out of the Einstein's theory of general
relativity (see \cite{kk, noll} for a complete list of references).
Recently there has been a renewed interest in black holes arising from
higher curvature corrections to Einstein-Hilbert action, partly due to
new developments in string theory \cite{asen, moura}. Low energy
limits of string theories give rise to effective models of gravity in
higher dimensions. These effective models of gravity involve higher
powers of the Riemann curvature tensor in the action in addition to
the usual Einstein-Hilbert term \cite{scherk}.  Among these higher
powers of Riemann curvature, the Gauss-Bonnet combination is of
special interest \cite{zwei,deser,wheeler, wheeler1, wiltsh}.  The
Gauss-Bonnet black holes has also been discussed in the context of
brane world models \cite{fromross}. The entropy and related
thermodynamic properties of GB black holes have been
discussed in \cite{ross, nojiri, cho, neu, cai, clunan}. Recently
GB black holes have received attention in the context of
possible production at the LHC \cite{alex}.

Scalar QN frequencies of the Gauss-Bonnet black holes were first
discussed long time ago \cite{vish1}, where the scalar modes of the
uncharged GB black hole for a few values of the Gauss-Bonnet coupling
were evaluated in $5$ and $6$ dimensions. Recently, in \cite{konoplya}
the QNMs for scalar perturbations of neutral and charged Gauss-Bonnet
black holes have been discussed. In a related work a detailed analysis
of scalar perturbations in the background of Gauss-Bonnet,
Gauss-Bonnet-de Sitter and Gauss-Bonnet-Anti-de Sitter black hole
space-time were presented \cite{kon1}. Using both numerical as well as
a semi-analytic WKB approach it has been shown that the scalar field
evolution is dependent on the cosmological constant and the GB
coupling. The late time behaviors of the scalar fields were also
discussed there.

In this paper we shall discuss the QNMs for tensor and vector
perturbations of the neutral Gauss-Bonnet black hole using a
semi-analytic WKB approach \cite{iyer}. It was claimed in the
literature \cite{konoplya} that the QN frequencies of the scalar field
perturbations and the tensor perturbations would be the same even in
the case of black holes arising in the Gauss-Bonnet gravity. We
explicitly show that they do not give rise to the same QN frequency
for GB black holes. We use the potential for tensor and vector mode
perturbations of static spherically symmetric solutions of the
Einstein equations with a Gauss-Bonnet term in dimensions $d>4$ as
analyzed in \cite{dotti, dottivector} respectively. Using the form of
the potential in \cite{dotti, dottivector} we obtain QN frequencies of
the Gauss-Bonnet black holes for different multipole numbers $l$ and
different values of the GB coupling. We have found that the
quasinormal behavior for the tensor and vector perturbation is
dependent on the GB coupling.  We have also shown that the QN
frequencies approach their Schwarzschild values when the GB coupling
tends to zero. The new results obtained in this paper are as follows:
\\
-The QN frequencies for vector and tensor perturbations of the
GB black hole are obtained using the WKB method.\\
-We show that the quasinormal frequencies for scalar field
perturbations and the tensor gravitational perturbations do not match
as was claimed in the literature \cite{konoplya} and the difference in
the result increases when the GB coupling is increased.\\
-It is shown that the Schwarzschild QN frequencies can be obtained
from the values of QN frequencies of the Gauss-Bonnet black hole when
the GB coupling goes to zero.

This paper is organized as follows: In Section 2 we shall briefly
review the Gauss-Bonnet action and the tensor perturbation of the
spherically symmetric Gauss-Bonnet space-time and will apply the third
order WKB formula to obtain QN frequencies for tensor and vector
perturbations of the Gauss-Bonnet black hole. Section 3 concludes
the paper with some discussion of our results and an outlook.

\sxn {Tensor and vector perturbation of spherically symmetric Gauss-Bonnet space-time and the Gauss-Bonnet black hole}

In space-time dimensions $d \geq 5$ the Einstein-Gauss-Bonnet action is given by
\be 
I=\frac{1}{16\pi G_d} \int d^dx \sqrt{-g} R + \alpha^{\prime}\int
  d^dx
  \sqrt{-g}(R_{\mu\nu\beta\gamma}R^{\mu\nu\beta\gamma}-4R_{\beta\gamma}R^{\beta\gamma}+R^2)  
,\label{alpha} 
\ee
where $G_d$ is the $d$-dimensional Newton's constant and the parameter
$\alpha^{\prime}$ denotes the Gauss-Bonnet coupling. We will choose
$G_d=1$ from now on and will consider only positive $\alpha^{\prime}$
which is consistent with the string expansion \cite{deser}.

We consider the metric in space-time dimension $d$ \cite{dotti}
\be
ds^2=-f(r)dt^2+f^{-1}(r)dr^2+r^2\bar g_{ij}dy^idy^j \label{metric}
\ee
where $\bar g_{ij}dy^i dy^j$ is the line element of the unit
$(d-2)$-dimensional sphere. Here, the latin indices $i, j, k,..$ and a
bar is used to denote tensors and operators in $S^{d-2}$\cite{dotti}. Considering the tensor perturbations
$g_{\mu\nu}\to g_{\mu\nu}+h_{\mu\nu}$, and using the following ansatz:
\be
h_{ij}(t,r,y)=r^2\phi(r,t)\bar h_{ij}(y)\,\,\,\rm{and}\,\,\, \phi(r,t)=e^{\omega t}\chi(r),
\ee
one obtains the linearized Einstein-Gauss-Bonnet equations \cite{dotti}
\be
\delta G_{(1)\mu}~^{\nu}+\alpha'\delta G_{(2)\mu}~^{\nu}=0\label{egb}
\ee
where
\be
G_{(2)\mu}~^{\nu}=R_{\sigma\mu}^{~\beta\gamma}R_{\beta\gamma}^{~\sigma\nu}
- 2R_{\beta}^{~\sigma}R_{\sigma\mu}^{~\beta\nu} -
2R_{\mu}^{~\sigma}R_{\sigma}^{~\nu} + RR_{\mu}^{~\nu} -
\frac{1}{4}\delta_{\mu}^{\nu}(R_{\sigma\beta}^{~\gamma\delta}R_{\gamma\delta}^{~\sigma\beta} -
4R_{\sigma}^{~\beta}R_{\beta}^{~\sigma} + R^2)
\ee 
From here on we will use subscript $T$ and $V$ to denote
quantities relating to tensor and vector perturbations respectively.
Now, introducing $\Phi(r)=\chi(r)K_T(r)$, Eqn. (\ref{egb})
reduces to a second order ordinary differential equation \cite{dotti}:
\be
-\frac{d^2\Phi}{dx^2}+V(r(x))\Phi=\omega^2\Phi\label{scheq}
\ee
Where $x$ is the ``tortoise coordinate'' defined by
$\frac{dx}{dr}=\frac{1}{f(r)}$, where $r=\infty$ corresponds to
$x=\infty$ and the event horizon $r=r_h$ corresponds to $x=-\infty$.
Now, $K_T(r)$ is defined as :
\be
K_T(r)  =  r^{\frac{d - 4}{2}}\sqrt{r^2~ + ~\alpha'(d - 4)\left[(d
    - 5)(1 - f(r))~ - ~r\frac{df}{dr}\right]}. \label{kr}
\ee
Now, the potential has the form \cite{dotti}
\be
V_T(r) = q_T(r) + \left(f\frac{d}{dr}ln(K_T)\right)^2 + f\frac{d}{dr}\left(f\frac{d}{dr}ln(K_T)\right)\label{potential}
\ee
with $q_T(r)$ being given by: 
\be
q_T(r)  =  \left(\frac{f(2 - \gamma)}{r^2}\right)\left(\frac{(1 - \alpha'
    f^{\prime\prime}(r))r^2 + \alpha'(d - 5)[(d - 6)(1 - f(r)) -
    2rf^{\prime}(r)]}{r^2 + \alpha'(d - 4)[(d - 5)(1 - f(r)) -
    rf^{\prime}(r)]}\right) \label{qr},
\ee
where $\gamma=-l(l+d-3)+2$ and $l=2, 3, 4, \cdots$ \cite{higuchi, rubin}. It is important to
mention that the results of Dotti and Gleiser readily reproduces the
results of Einstein gravity in the $\alpha' \to 0$ limit, which was
studied by Ishibashi and Kodama \cite{ishi, ishi1}.

The vector perturbations of the Einstein-Gauss-Bonnet spacetime was
discussed in \cite{dottivector}. The method used there is  different
from the computations carried out for the tensor
perturbations. Without going into the details of the techniques used
in \cite{dottivector}, we mention here the potential for vector
perturbation:
\be
V_V(r)=q_V(r)+\frac{1}{K_V(r)}\left[\left(\frac{d^2K_V}{dr^2}\right)f^2+\left(\frac{dK_V}{dr}\right)f\frac{df}{dr}\right]\label{VV}
\ee
where
\be
K_V(r)=\left[r^{(d-2)}+\alpha^{\prime}(d-4)r^2\frac{\partial}{\partial r}(r^{(d-5)}(1-f))\right]
\ee
and
\be
q_V(r)=\frac{f[(l(l+d-3)-1)-(d-3)]H}{r^2},\,\,l=1,2,3...
\ee
where,
\bea
H=\frac{\alpha^{\prime 2}(d-2)(d-4)^2(d-5)(d-3)\psi^2+2\alpha^{\prime}(d-2)(d-4)(d-5)\psi+4(d-4)\Lambda\alpha^{\prime}+2(d-2)}{2(d-2)((d-3)(d-4)\psi\alpha^{\prime}+1)^2}\nonumber
\eea
Here $\psi(r)$ is a solution of 
\be
\frac{(d-2)(d-3)(d-4)}{4}\psi^2+\frac{(d-2)}{2\alpha^{\prime}}\psi-\frac{\Lambda}{(d-1)\alpha^{\prime}}=\frac{M}{\alpha^{\prime}r^{(d-1)}}
\ee
Where $\Lambda$ is the cosmological constant. In what follows, we
shall take $\Lambda=0$.\\

{\large{\bf 2a. Uncharged Gauss-Bonnet Black Hole}}\\

The metric for spherically symmetric asymptotically flat Gauss-Bonnet
black hole solution of mass $M$ is given by Eqn. (\ref{metric}), where
$f(r)$ has the form \cite{deser}
\be
f(r) = 1 + \frac{r^2}{2\alpha} -
\frac{r^2}{2\alpha}\sqrt{1+\frac{8\alpha M}{r^{d-1}}}, \label{fr}
\ee
where, 
\be
\alpha=16\pi G_d (d-3)(d-4)\alpha'.
\ee
For $\alpha'>0$, this black hole admits only a single horizon
\cite{deser}. The horizon $r=r_h$ is determined by the real positive
solution of the equation
\be
r_h^{d-3}+\alpha r_h^{d-5}=2M. \label{horizon}
\ee

For most spacetime geometries, the wave equation governing the QNMs is
not exactly solvable. Various numerical schemes have been used in the
literature to find the QN frequencies, which include direct
integration of the wave equation in the frequency domain
\cite{chandra}, P\"{o}schl-Teller approximation \cite{ferrari}, WKB
method \cite{will,will2,iyer,kon2}, phase integral method
\cite{Andersson, and1} and continued fraction method \cite{leaver}. We
will use the WKB method in this paper.

The formula for QN frequencies using third order WKB approach is given
by \cite{will,iyer}
\be
\omega^2=[V_0+(-2V_0^{\prime\prime})^{1/2}\tilde\Lambda(n)]-i(n+\frac{1}{2})(-2V_0^{\prime\prime})^{1/2}[1+\tilde\Omega(n)].\label{freq}
\ee
where, $\tilde\Lambda=\Lambda/i$ and
$\tilde\Omega=\Omega/(n+\frac{1}{2})$ and $\Lambda$ and $\Omega$ are
given by
\bea
\Lambda(n)&=&\frac{i}{(-2V^{\prime\prime}_0)^{1/2}}\left[\frac{1}{8}\left(\frac{V^{(4)}_0}{V^{\prime\prime}_0}\right)
\left(\frac{1}{4}+\nu^2\right)-\frac{1}{288}\left(\frac{V^{(3)}_0}{V^{\prime\prime}_0}\right)^2
(7+60\nu^2)\right],\nonumber\\
\Omega(n)&=&\frac{(n+\frac{1}{2})}{(-2V^{\prime\prime}_0)^{1/2}}\bigg [\frac{5}{6912}
\left(\frac{V^{(3)}_0}{V^{\prime\prime}_0}\right)^4
(77+188\nu^2)\nonumber\\&&-
\frac{1}{384}\left(\frac{V^{(3)^2}_0V^{(4)}_0}{V^{\prime\prime^3}_0}\right)
(51+100\nu^2)
+\frac{1}{2304}\left(\frac{V^{(4)}_0}{V^{\prime\prime}_0}\right)^2(67+68\nu^2)
\nonumber\\&&+\frac{1}{288}
\left(\frac{V^{(3)}_0V^{(5)}_0}{V^{\prime\prime^2}_0}\right)(19+28\nu^2)-\frac{1}{288}
\left(\frac{V^{(6)}_0}{V^{\prime\prime}_0}\right)(5+4\nu^2)\bigg ].\label{wl}
\eea
Where, $V^{(n)}_0=(d^nV/dx^n)_{x=x_0}$ and $\nu=n+1/2$.

It may be mentioned here that the accuracy of the WKB method depends
on the multipole number $l$ and the overtone number $n$. It has been
shown \cite{cardosoyousi} that the WKB approach is a good one for
$l>n$, i.e. the numerical and the WKB results are in good agreement if
$l>n$, but the WKB approach is not so good if $l=n$ and not at all
applicable for $l<n$.

Now, using Eqns. (\ref{freq}), (\ref{wl}) and Eqns.
(\ref{potential}), (\ref{VV}) we numerically evaluate the QNMs of
this black hole.  In Tables 1-3 we present the quasinormal frequencies
for the uncharged Gauss-Bonnet black hole for tensor perturbation,
where the metric is given by Eqn.  (\ref{metric}) with $f(r)$ being
given by Eqn. (\ref{fr}) in $d=5, 7 \,\, \rm{and}\,\, 8$ for $l = 2, 3
\,\,\rm{and}\,\, 4$. Here the black hole in $d=6$ is not considered
because of the fact that it is unstable to tensor mode perturbations
\cite{dotti}. Note also that in $d=5$, the horizon is at
$r_h=\sqrt{2-\alpha}$. Thus for $d=5$, black hole type solutions exist
only for $\alpha<2$ and hence the corresponding entries in the table
are kept empty. We have calculated the frequencies for fundamental
overtone $(n=0)$, which will be dominating in a signal. We have set
$M=1$ in our calculations.

The behavior of the real part of the quasinormal mode frequencies with
the Gauss Bonnet coupling $\alpha$ in seven and eight dimensions
respectively are shown in Figure \ref{fig2} and \ref{fig3}. In
Figure \ref{fig4} and \ref{fig5}, we have plotted the behavior of
imaginary part of the quasinormal frequency with the coupling $\alpha$
in $d=7 \,\rm{and}\, 8$ respectively.  It is observed that the
real part of the quasinormal mode frequency increases with $\alpha$,
i.e the real oscillation frequency increases as the Gauss-Bonnet
coupling increases. But the imaginary part shows a different kind of
behavior, the damping decreases as $\alpha$ is increased from some
lower values until some minimum value and then the damping increases
as we go to larger values of $\alpha$. 

In \cite{konoplya}, scalar perturbations of the Gauss-Bonnet black
hole were considered and it was shown that the quasinormal modes have
greater oscillation frequency and greater damping rate at large
$\alpha$. However, at moderate $\alpha$, the damping rate is
decreasing until some minimum value and then begin to grow as a
function of $\alpha$. It has been mentioned in \cite{konoplya} that
the tensor type gravitational potential and the scalar field potential
coincide as in the case of higher dimensional black holes in Einstein
gravity as was established in \cite{ishi}. But it was shown by
Dotti and Gleiser \cite{dotti} that the tensor perturbations of the
Einstein-Gauss-Bonnet black hole is completely different from that of
the tensor perturbations of black holes in Einstein gravity and as the
Gauss-Bonnet coupling $\alpha$ goes to zero, the results of Einstein
gravity can be recovered. We give a plot in fig. 1 to show how the
potential for tensor gravitational perturbations and the scaler
perturbation differ for a particular value of $\alpha$ and $l$.
\begin{figure}[here]
\begin{center}
\centerline{\hspace{0.3mm}
\rotatebox{0}{\epsfxsize=14cm\epsfbox{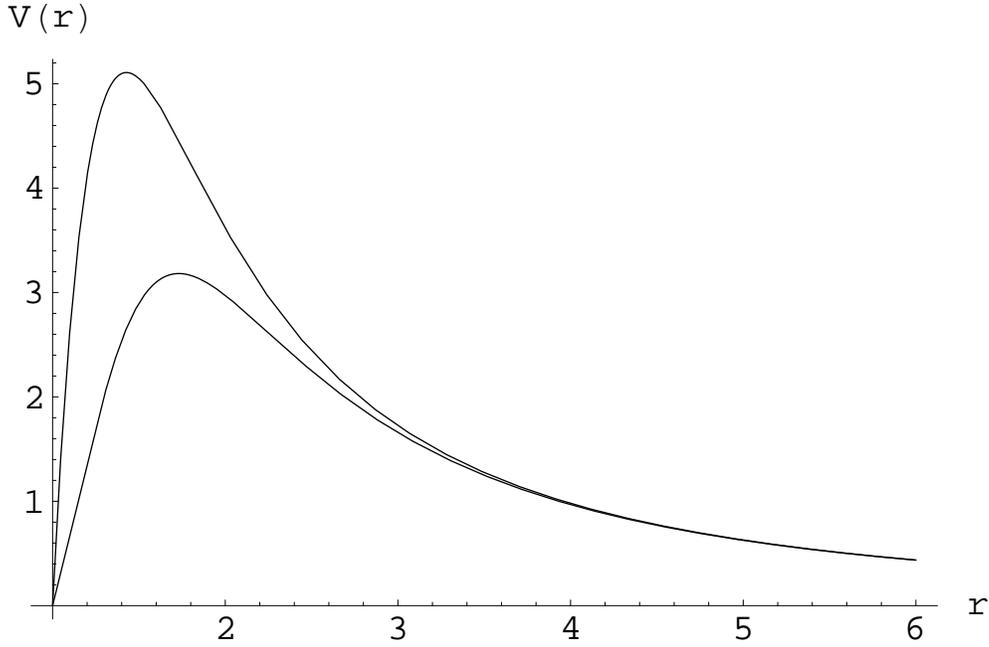}}}
\hspace{3.3cm}\caption[]{Comparison between the potential for tensor
  gravitational perturbations and scalar field perturbations for
  Gauss-Bonnet black hole for $\alpha=1$ and $l=2$. The lower curve
  denotes the potential for tensor gravitational perturbations while
  the upper one denotes the same for scaler field perturbations.}
\protect\label{fig1}
\end{center}
\end{figure}
Here we have used the potential for tensor perturbations obtained by
Dotti and Gleiser and derived the QN frequencies for tensor
perturbations of the Gauss-Bonnet black holes.  The qualitative
behavior of real and imaginary part of QN frequencies for tensor
perturbations with $\alpha$ that we have obtained, is similar to that
obtained in \cite{konoplya}, where scalar perturbations were
considered. If we compare our results with the results obtained in
\cite{konoplya} for $d=7,\, \rm{and} \,\, l=2$, then we find that the
real and imaginary parts of the QN frequency are different and the
difference increases as $\alpha$ is increased. For example, the real
and imaginary parts of our result differs from \cite{konoplya} by
$0.5\%$ and $0.7\%$ respectively for $\alpha=0.1$, but the difference
is $9\%$ and $1\%$ respectively for $\alpha=20$.

Here, we have used third order WKB approximation to find the
quasinormal frequency of the Gauss-Bonnet black hole as we were unable
to use sixth order corrections to this formula due to computational
complcacies. However, there is not a very huge difference between
third and sixth order result of QN frequencies for $l>n$ \cite{kon2}.
In the case for $l=0$, where the lowest overtone implies $l=n$, the
WKB formula has a large amount of error when third order results are
compared with the accurate numerical result for the Schwarzschild
black hole, while the sixth order results almost matches with the
numerical result \cite{kon2}. But as we are considering tensor
perturbations, where the lowest overtone implies $l>n$, we expect that
the third order WKB treatment will be of good accuracy.

We have also verified that indeed the Schwarzschild QN frequency for
tensorial perturbations can be obtained if $\alpha$ tends to zero.
Now, to see that the numerical value of the QN frequency of the
Gauss-Bonnet black hole approaches the value of the quasinormal mode
frequency of the Schwarzschild black hole, we first consider six
sample modes ($d=5,\, l=2, \, n=0$), found out from different small
values of $\alpha$, $\alpha = 1/100, 1/200, 1/300, 1/400, 1/500$ as
considered in \cite{konoplya}.  Then we find the quadratic fit and get
\be 
\omega = (1.06566- 0.25305i) +
(0.06285+ 0.05782i)\alpha + (0.16704 - 0.06213i)\alpha^2 + O[\alpha^3]
\ee
 So, we observe that the fit is approaching the fundamental ($n=0$)
and $l=2$ quasinormal frequency for tensor type perturbation of the
Schwarzschild black hole in five dimension, which is $1.0681-0.2529 i$
\cite{konoplya3}. It may be mentioned here that in \cite{konoplya3},
sixth order WKB correction formula was used to determine the
quasinormal frequencies, but due to the complicacy of the potential
here we were forced to use third order WKB correction formula.
Therefore we expect that one can get more exact value of the frequency
from the fit if sixth order WKB correction formula is used.

The numerical results for the quasinormal modes for vector
perturbations are given in Tables 4-7. Here we have considered the
values of $l=1, 2, 3$ and $4$ in $d=5, 6, 7$ and $8$ dimensions. As
there is no problem of stability of the black hole in $d=6$ under
vector perturbations like the tensor perturbation case, we consider
$d=6$ also.  But the problem with $d=6$ is that we are getting double
peaks in the potential for larger values of $\alpha$. As that causes
problem in the WKB analysis we left the corresponding entries in the
tables blank. There is also some problem with the QN frequencies of
$l=1$ in the case of vector perturbations. The nature of the
dependence of QN frequencies for $l=1$ differ a little bit with those
of $l=2,\, 3 \,\rm{and}\, 4$. The probable cause for this might be due
to the fact that we are using third order WKB approach to determine
the QN frequencies. If sixth order treatment is used, we think that
results from $l=1$ will be less erroneous. The plot for real and
imaginary parts of the QN frequency in seven and eight dimensions are
shown in fig. 6 - fig. 9 respectively.

Like the tensorial perturbations, here also we checked that
Schwarzschild QN frequencies can be obtained from lower values of
$\alpha$. For that we first consider five sample modes ($d=5,\, l=2,
\, n=0$), found out from different small values of $\alpha$, $\alpha =
1/100, 1/200, 1/300, 1/400, 1/500$. Then we find the quadratic fit and get
\be 
\omega = (0.80559 - 0.23545i) + (0.03376 + 0.005496i)\alpha +
(0.00465 + 0.00729i)\alpha^2 + O[\alpha^3] 
\ee
We note that the fit is approaching the fundamental ($n=0$) and $l=2$
quasinormal frequency for vector type perturbation of the
Schwarzschild black hole in five dimension, which is $0.8056 -
0.2355i$ \cite{konoplya3}

\begin{table}[here]
\begin{center}
\begin{tabular}[b]{|c|c|c|c|}  \hline \hline
$\alpha$ & \multicolumn{3}{|c|}{$\omega$}\\\cline{2-4} & d=5 & d=7 &
d=8 \\
\hline
0.1  & 1.07234-0.24737i &  2.09158-0.51813i & 2.57245-0.63574i \\
0.2  & 1.07949-0.24175i &  2.09568-0.50563i & 2.57391-0.61982i \\
0.5  & 1.10393-0.22500i &  2.11523-0.47490i & 2.58863-0.58251i \\
1.0  & 1.15641-0.19529i &  2.16399-0.44120i & 2.63419-0.54656i \\
5.0  &        -         &  2.64723-0.33309i & 3.02722-0.42703i \\
10.0 &        -         &  3.31276-0.43274i & 3.43874-0.47634i \\
15.0 &        -         &  3.84348-0.52104i & 3.78370-0.53669i \\
20.0 &        -         &  4.23657-0.59334i & 4.05205-0.58548i \\ 
\hline\hline
\end{tabular}
\caption{Quasinormal mode frequencies for tensor perturbation for $l=2$ in $d=5, 7$ and $8$}
\end{center}
\end{table}
\begin{table}[!]
\begin{center}
\begin{tabular}[b]{|c|c|c|c|}  \hline\hline
$\alpha$ & \multicolumn{3}{|c|}{$\omega$}\\\cline{2-4} & d=5  & d=7 & d=8 \\ \hline\hline
0.1  & 1.42825-0.24648i &  2.62080-0.51715i &  3.15722-0.63605i \\
0.2  & 1.43818-0.24123i &  2.62803-0.50632i &  3.16206-0.62266i \\
0.5  & 1.47124-0.22497i &  2.65471-0.47757i &  3.18328-0.58829i \\
1.0  & 1.54074-0.19430i &  2.71209-0.44140i &  3.23448-0.54912i \\
5.0  &        -         &  3.33102-0.33905i &  3.70504-0.42783i \\
10.0 &        -         &  4.19705-0.43708i &  4.24267-0.48196i \\
15.0 &        -         &  4.87089-0.52488i &  4.67515-0.54131i \\
20.0 &        -         &  5.36533-0.59728i &  5.00776-0.58973i \\ 
\hline\hline
\end{tabular}
\caption{Quasinormal mode frequencies for tensor perturbation for $l=3$ in $d=5, 7$ and $8$}
\end{center}
\end{table}

\begin{table}[!]
\begin{center}
\begin{tabular}[b]{|c|c|c|c|}  \hline\hline
$\alpha$ & \multicolumn{3}{|c|}{$\omega$}\\\cline{2-4} & d=5  & d=7 & d=8 \\
 \hline\hline
0.1  & 1.78374-0.24604i &  3.14573-0.51602i & 3.73477-0.63489i \\
0.2  & 1.79630-0.24094i &  3.15560-0.50603i & 3.74238-0.62289i \\
0.5  & 1.83780-0.22488i &  3.18919-0.47854i & 3.77012-0.59069i \\
1.0  & 1.92439-0.19367i &  3.25649-0.44120i & 3.82896-0.55030i \\
5.0  &        -         &  4.01271-0.34219i & 4.38417-0.42912i \\
10.0 &        -         &  5.07277-0.43935i & 5.04128-0.48505i \\
15.0 &        -         &  5.88797-0.52690i & 5.55942-0.54389i \\
20.0 &        -         &  6.48321-0.59934i & 5.95548-0.59212i \\ 
\hline\hline
\end{tabular}
\caption{Quasinormal mode frequencies for tensor perturbation for $l=4$ in $d=5, 7$ and $8$} 
\end{center}
\end{table}

\begin{figure}[!]
\begin{minipage}[t]{8cm}
\vspace{-10pt}
\centerline{\hspace{6.3mm}
\rotatebox{-90}{\epsfxsize=8cm\epsfbox{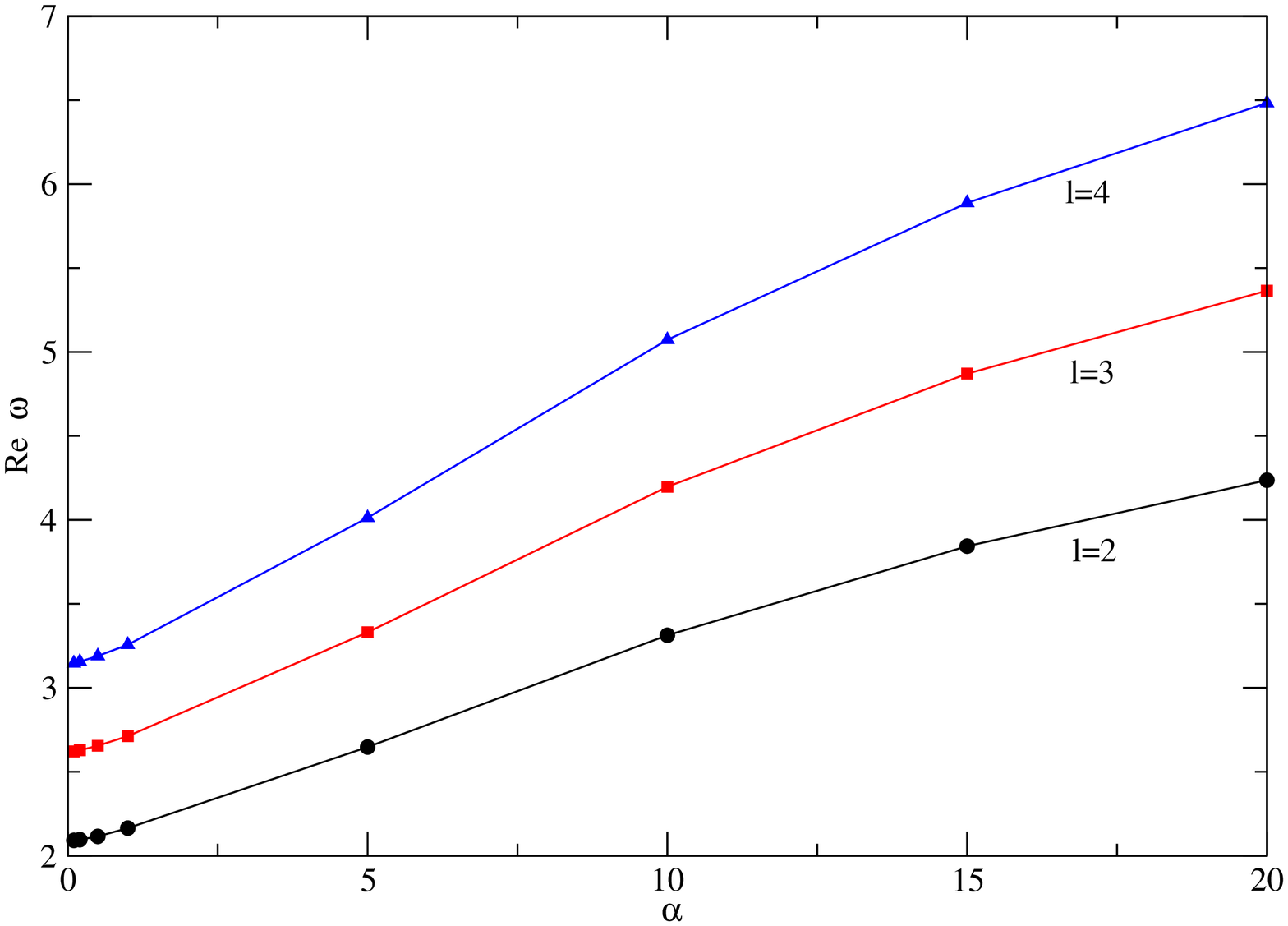}}}
\hspace{3.3cm}\caption[]{Re $\omega$ as a function of Gauss-Bonnet
  coupling $\alpha$ for $l=2, 3, 4$ and $n=0$ in $d=7$ for tensor
  perturbations.} 
\protect\label{fig2}
\end{minipage}
\hfill
\begin{minipage}[t]{8cm}
\vspace{-10pt}
\centerline{\hspace{11.3mm}
\rotatebox{-90}{\epsfxsize=8cm\epsfbox{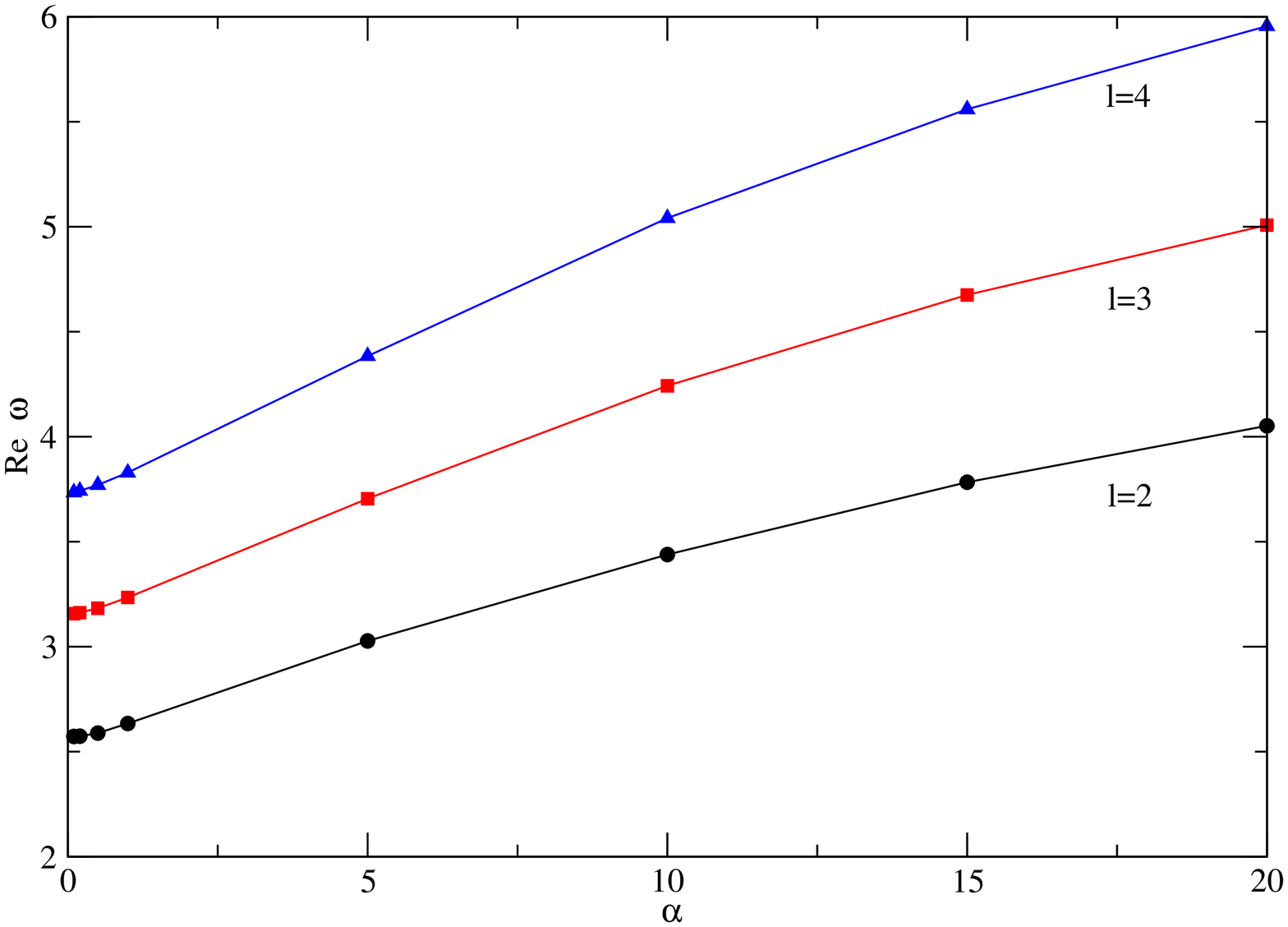}}}
\hspace{3.3cm}\caption[]{Re $\omega$ as a function of Gauss-Bonnet
  coupling $\alpha$ for $l=2, 3, 4$ and $n=0$ in $d=8$ for tensor
  perturbations.}
\protect\label{fig3}
\end{minipage}
\end{figure}

\begin{figure}[!]
\begin{minipage}[t]{8cm}
\vspace{-10pt}
\centerline{\hspace{6.3mm}
\rotatebox{-90}{\epsfxsize=8cm\epsfbox{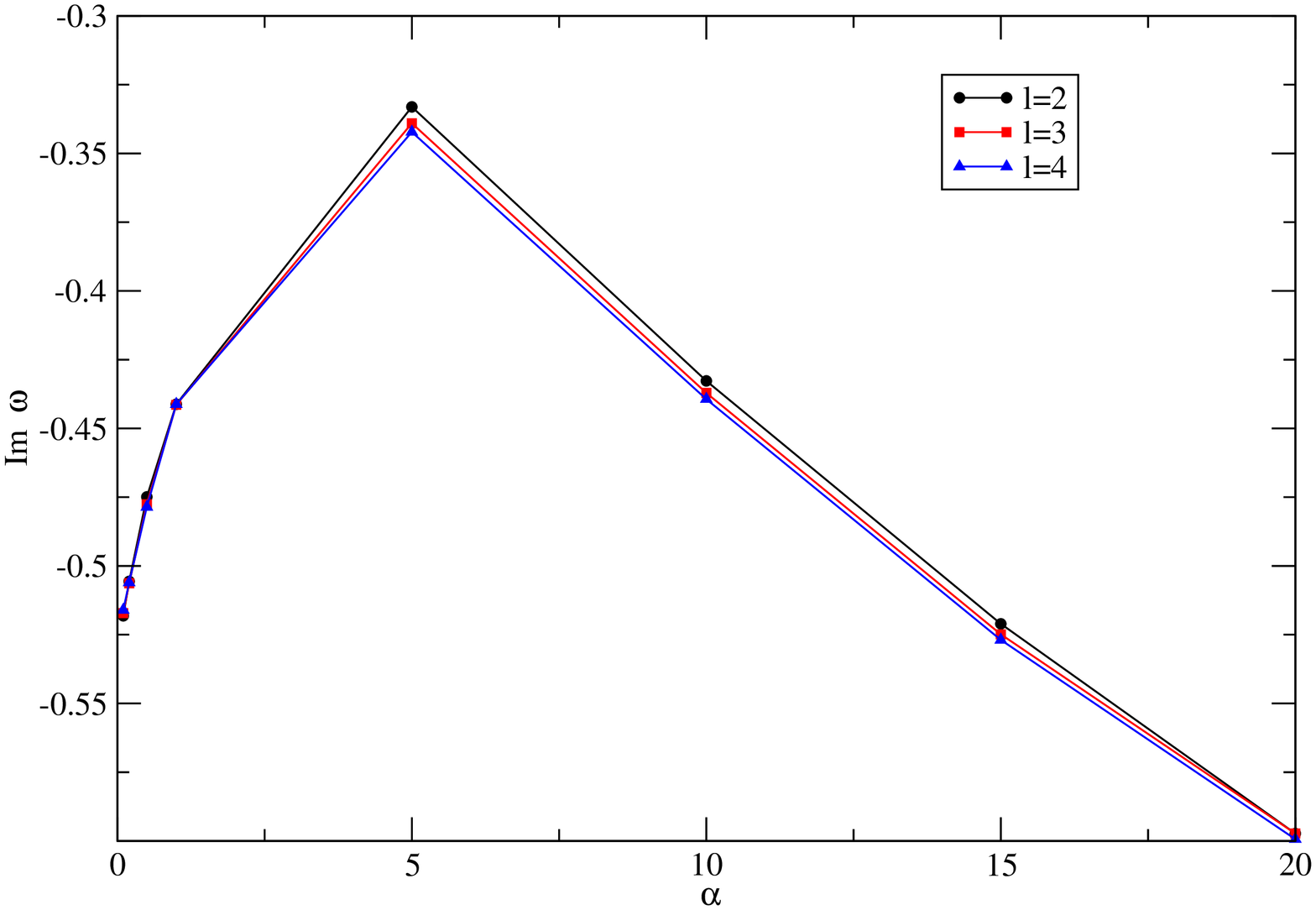}}}
\hspace{3.3cm}\caption[]{Im $\omega$ as a function of Gauss-Bonnet
  coupling $\alpha$ for $l=2, 3, 4$ and $n=0$ in $d=7$ for tensor
  perturbations.}
\protect\label{fig4}
\end{minipage}
\hfill
\begin{minipage}[t]{8cm}
\vspace{-10pt}
\centerline{\hspace{11.3mm}
\rotatebox{-90}{\epsfxsize=8cm\epsfbox{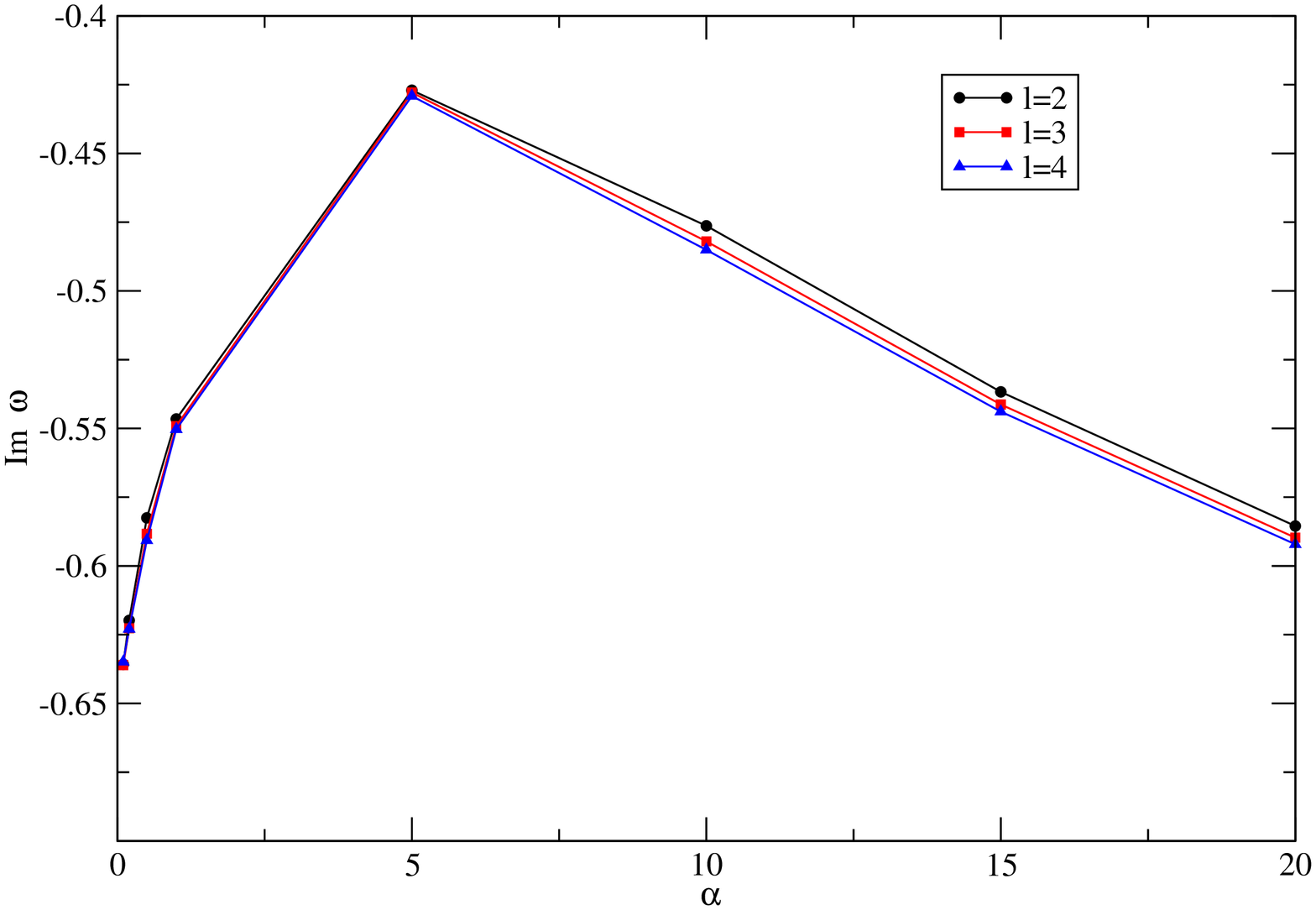}}}
\hspace{3.3cm}\caption[]{Im $\omega$ as a function of Gauss-Bonnet
  coupling $\alpha$ for $l=2, 3, 4$ and $n=0$ in $d=8$ for tensor
  perturbations.}
\protect\label{fig5}
\end{minipage}
\end{figure}

\begin{table}[!]
\begin{center}
\begin{tabular}[b]{|c|c|c|c|c|}  \hline \hline
$\alpha$ & \multicolumn{4}{|c|}{$\omega$}\\\cline{2-5} & d=5 & d=6 &
d=7 & d=8 \\
\hline
0.1  & 0.38852-0.23491i & 0.71811-0.37339i & 1.08031-0.50041i & 1.46373-0.61715i \\
0.2  & 0.38845-0.23310i & 0.71772-0.37044i & 1.07951-0.49660i & 1.46250-0.61271i \\
0.5  & 0.38801-0.22691i & 0.71640-0.36036i & 1.07711-0.48362i & 1.45897-0.59754i \\
1.0  & 0.38612-0.21348i & 0.71389-0.33903i & 1.07361-0.45689i & 1.45410-0.56680i \\
5.0  &        -         & 0.95379-0.07595i & 1.33952-0.45084i & 1.66619-0.49783i \\
10.0 &        -         &        -         & 1.46092-0.36981i & 1.79779-0.39821i \\
15.0 &        -         &        -         & 1.68586-0.47334i & 1.94080-0.49012i \\
20.0 &        -         &        -         & 1.87237-0.54573i & 2.07360-0.54533i \\ 
\hline\hline
\end{tabular}
\caption{Quasinormal mode frequencies for vector perturbation for
  $l=1$ in $d=5, 6, 7$ and $8$}
\end{center}
\end{table}
\begin{table}[!]
\begin{center}
\begin{tabular}[b]{|c|c|c|c|c|}  \hline \hline
$\alpha$ & \multicolumn{4}{|c|}{$\omega$}\\\cline{2-5} & d=5 & d=6 &
d=7 & d=8 \\
\hline
0.1  & 0.80908-0.22980i & 1.22692-0.37114i & 1.64991-0.50074i & 2.07904-0.61909i \\
0.2  & 0.81282-0.22383i & 1.22918-0.36377i & 1.65098-0.49298i & 2.07920-0.61118i \\
0.5  & 0.82668-0.20401i & 1.23772-0.33942i & 1.65571-0.46751i & 2.08109-0.58516i \\
1.0  & 0.86280-0.16695i & 1.26100-0.29391i & 1.67071-0.42051i & 2.09037-0.53716i \\
5.0  &        -         & 1.79853-0.29350i & 2.03929-0.32006i & 2.41661-0.43291i \\
10.0 &        -         &        -         & 2.37045-0.45111i & 2.66779-0.49225i \\
15.0 &        -         &        -         & 2.53501-0.45766i & 2.83381-0.55573i \\
20.0 &        -         &        -         & 2.62139-0.34334i & 2.92724-0.55988i \\ 
\hline\hline
\end{tabular}
\caption{Quasinormal mode frequencies for vector perturbation for
  $l=2$ in $d=5, 6, 7$ and $8$}
\end{center}
\end{table}
\begin{table}[!]
\begin{center}
\begin{tabular}[b]{|c|c|c|c|c|}  \hline \hline
$\alpha$ & \multicolumn{4}{|c|}{$\omega$}\\\cline{2-5} & d=5 & d=6 &
d=7 & d=8 \\
\hline
0.1  & 1.22732-0.23143i & 1.74514-0.36568i & 2.22978-0.49349i & 2.70311-0.61275i \\
0.2  & 1.23556-0.22612i & 1.75199-0.35737i & 2.23449-0.48374i & 2.70603-0.60251i  \\
0.5  & 1.26382-0.20983i & 1.77616-0.33261i & 2.25186-0.45407i & 2.71766-0.57078i \\
1.0  & 1.32628-0.18150i & 1.83089-0.29695i & 2.29356-0.40943i & 2.74804-0.52034i \\
5.0  &        -         & 2.60496-0.31368i & 2.80530-0.34825i & 3.15756-0.41819i \\
10.0 &        -         &        -         & 3.23719-0.47697i & 3.51637-0.51655i  \\
15.0 &        -         &        -         & 3.40873-0.51161i & 3.70743-0.58013i \\
20.0 &        -         &        -         & 3.46675-0.48649i & 3.80401-0.59800i \\ 
\hline\hline
\end{tabular}
\caption{Quasinormal mode frequencies for vector perturbation for
  $l=3$ in $d=5, 6, 7$ and $8$}
\end{center}
\end{table}
\begin{table}[!]
\begin{center}
\begin{tabular}[b]{|c|c|c|c|c|}  \hline \hline
$\alpha$ & \multicolumn{4}{|c|}{$\omega$}\\\cline{2-5} & d=5 & d=6 &
d=7 & d=8 \\
\hline
0.1  & 1.62271-0.23563i & 2.25087-0.36801i & 2.80549-0.49205i & 3.32693-0.60901i \\
0.2  & 1.63381-0.23061i & 2.26119-0.36020i & 2.81365-0.48223i & 3.33289-0.59806i \\
0.5  & 1.67107-0.21492i & 2.29598-0.33755i & 2.84194-0.45386i & 3.35437-0.56581i \\
1.0  & 1.75070-0.18540i & 2.36771-0.30442i & 2.90198-0.41485i & 3.40253-0.51999i \\
5.0  &        -         & 3.35548-0.32099i & 3.54531-0.35999i & 3.90098-0.42763i \\
10.0 &        -         &        -         & 4.06418-0.48967i & 4.34217-0.52963i \\
15.0 &        -         &        -         & 4.24763-0.53548i & 4.55712-0.59556i \\
20.0 &        -         &        -         & 4.30050-0.53964i & 4.65663-0.62039i \\ 
\hline\hline
\end{tabular}
\caption{Quasinormal mode frequencies for vector perturbation for
  $l=4$ in $d=5, 6, 7$ and $8$}
\end{center}
\end{table}

\begin{figure}[!]
\begin{minipage}[t]{8cm}
\vspace{-10pt}
\centerline{\hspace{6.3mm}
\rotatebox{-90}{\epsfxsize=8cm\epsfbox{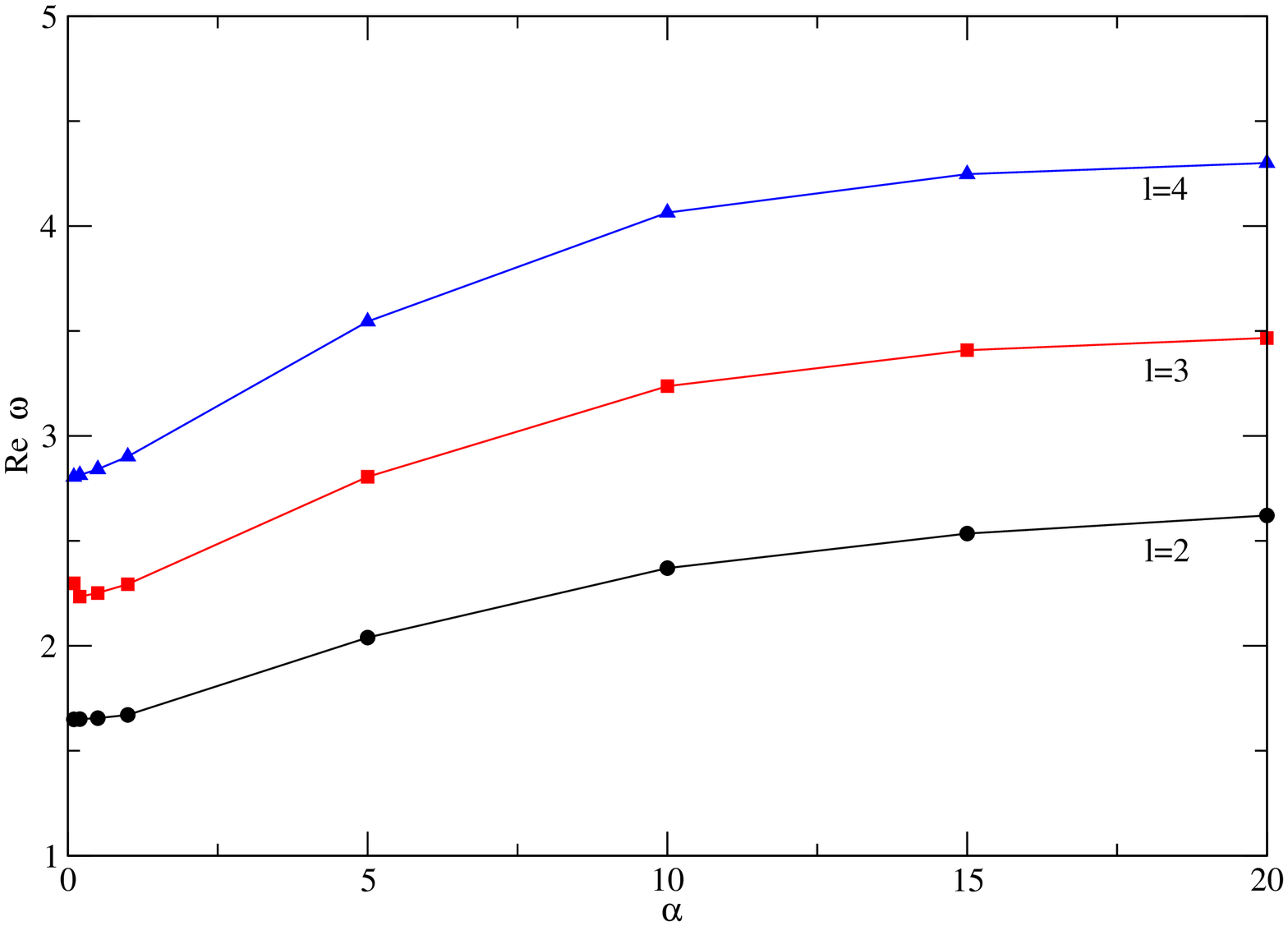}}}
\hspace{3.3cm}\caption[]{Re $\omega$ as a function of Gauss-Bonnet
  coupling $\alpha$ for $l=2, 3, 4$ and $n=0$ in $d=7$ for vector perturbations.}
\protect\label{fig6}
\end{minipage}
\hfill
\begin{minipage}[t]{8cm}
\vspace{-10pt}
\centerline{\hspace{11.3mm}
\rotatebox{-90}{\epsfxsize=8cm\epsfbox{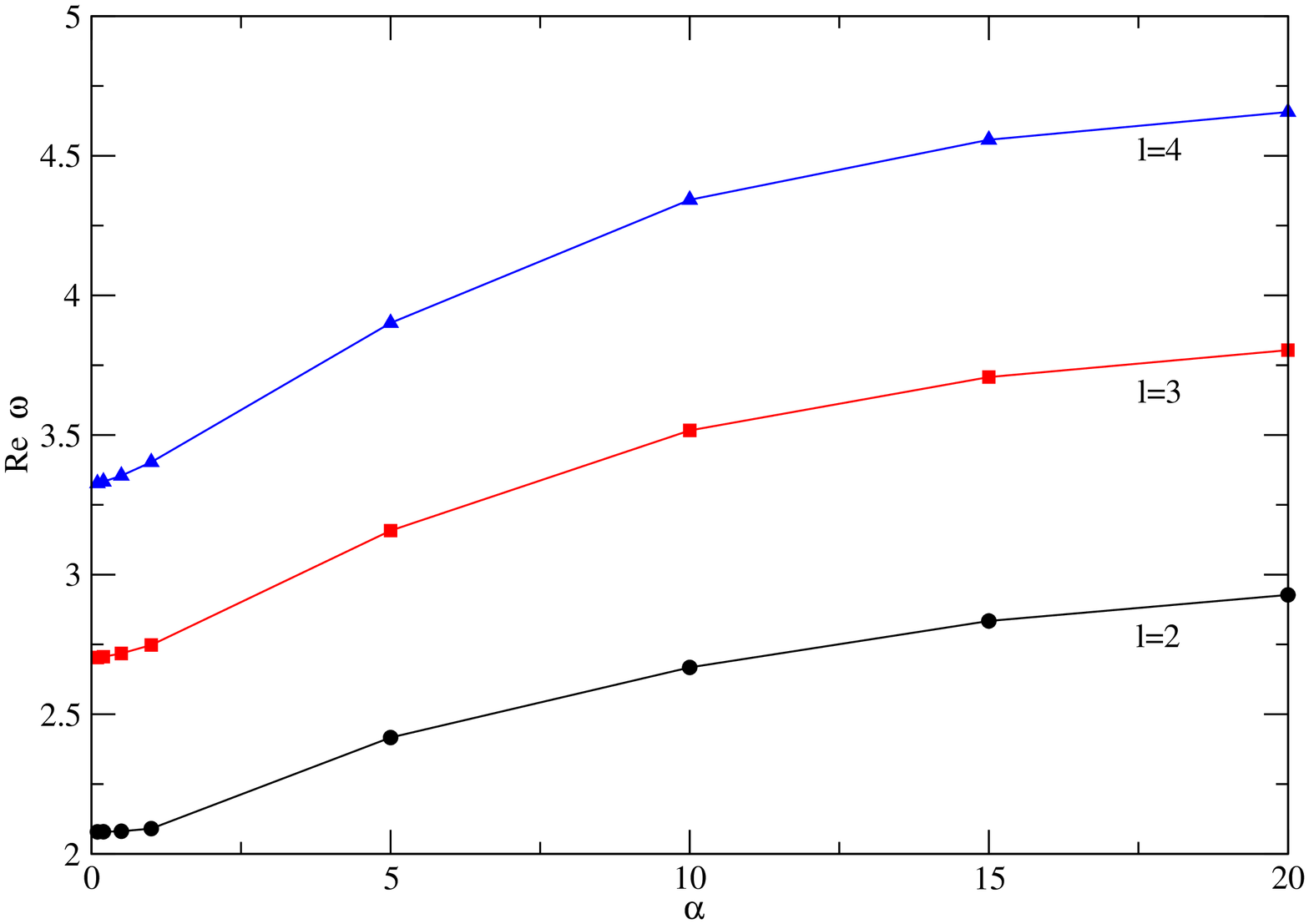}}}
\hspace{3.3cm}\caption[]{Re $\omega$ as a function of Gauss-Bonnet
  coupling $\alpha$ for $l=2, 3, 4$ and $n=0$ in $d=8$ for vector perturbations.}
\protect\label{fig7}
\end{minipage}
\end{figure}
\begin{figure}[!]
\begin{minipage}[t]{8cm}
\vspace{-10pt}
\centerline{\hspace{6.3mm}
\rotatebox{-90}{\epsfxsize=8cm\epsfbox{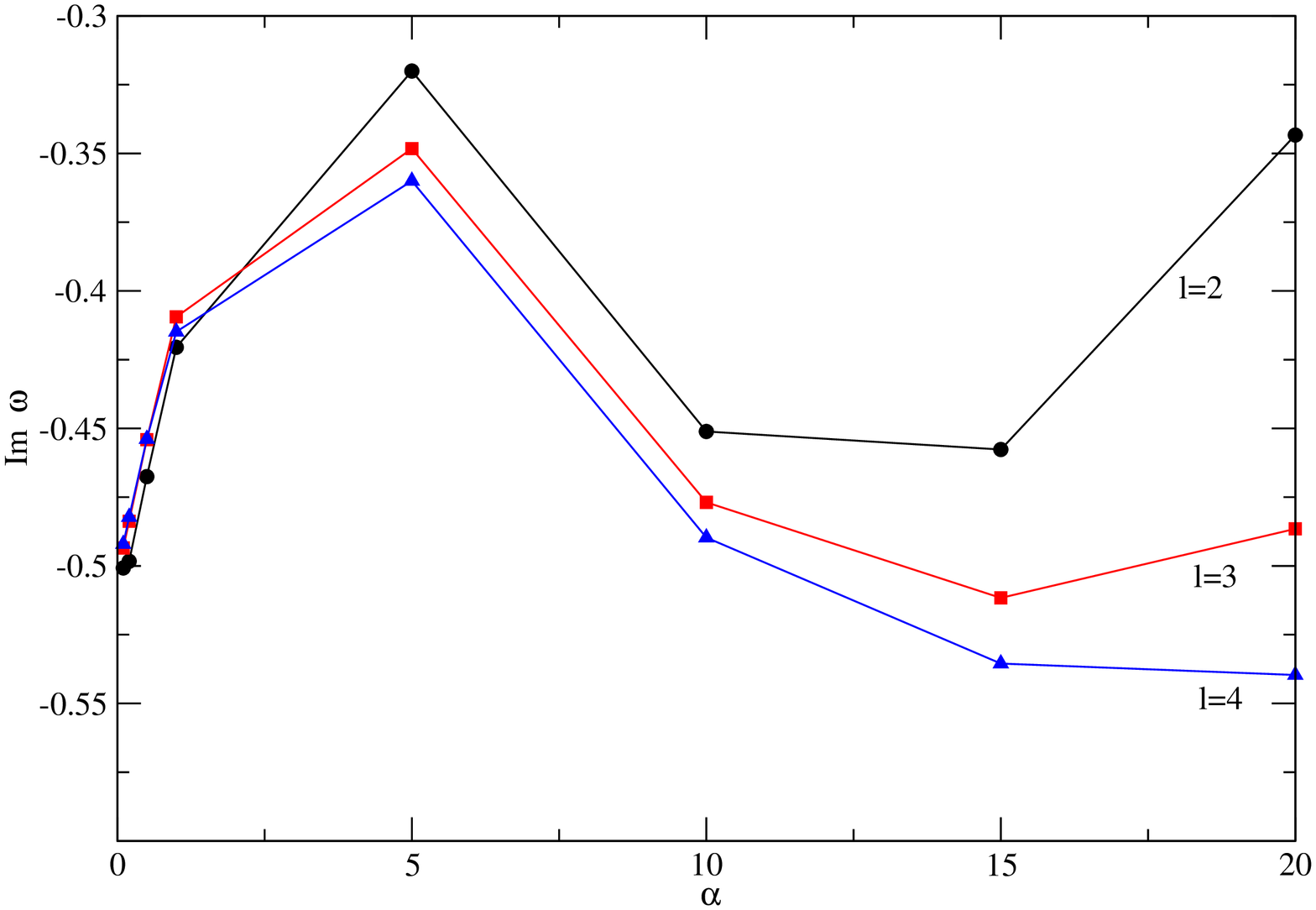}}}
\hspace{3.3cm}\caption[]{Im $\omega$ as a function of Gauss-Bonnet
  coupling $\alpha$ for $l=2, 3, 4$ and $n=0$ in $d=7$ for vector perturbations.}
\protect\label{fig8}
\end{minipage}
\hfill
\begin{minipage}[t]{8cm}
\vspace{-10pt}
\centerline{\hspace{11.3mm}
\rotatebox{-90}{\epsfxsize=8cm\epsfbox{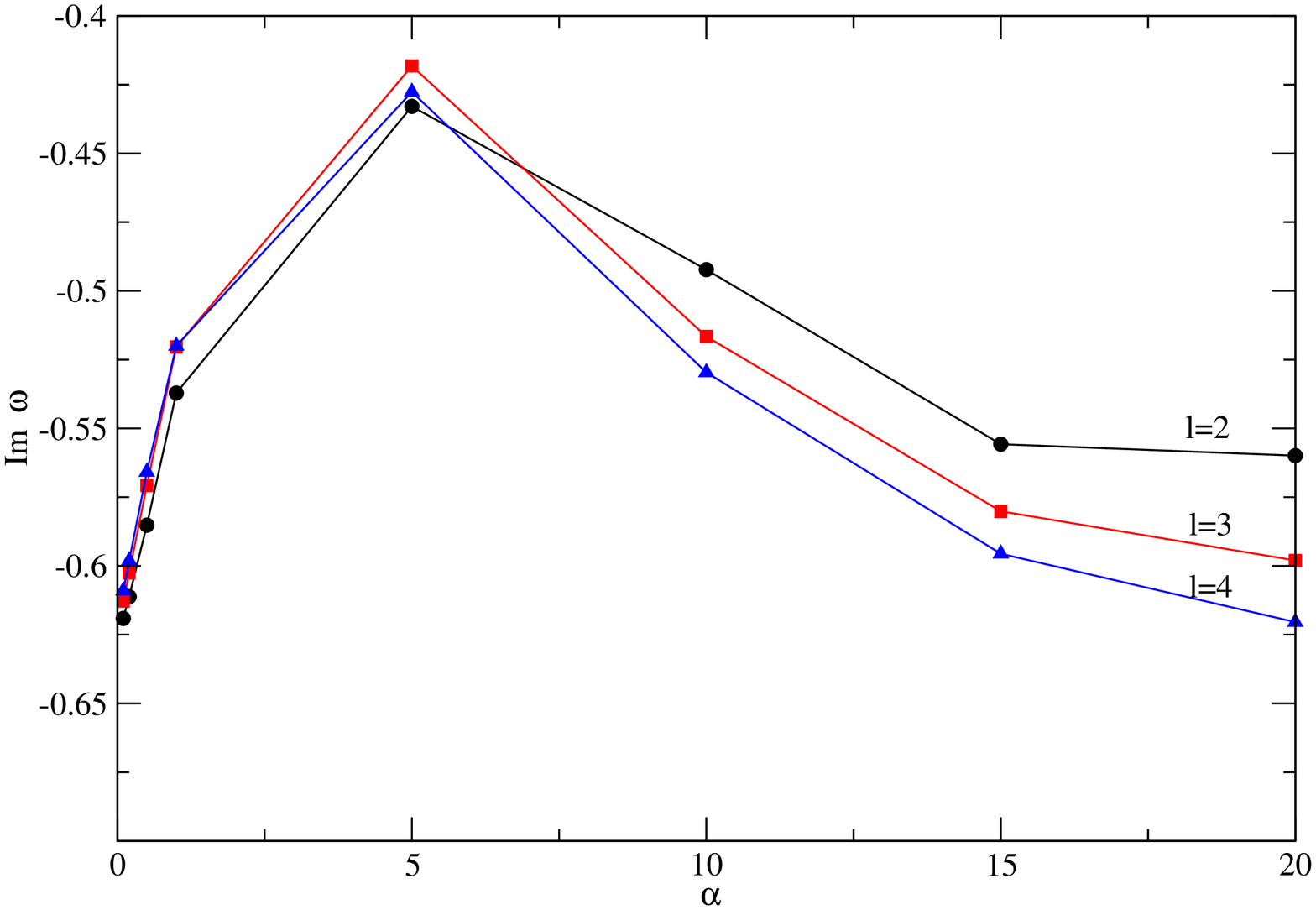}}}
\hspace{3.3cm}\caption[]{Im $\omega$ as a function of Gauss-Bonnet
  coupling $\alpha$ for $l=2, 3, 4$ and $n=0$ in $d=8$ for vector perturbations.}
\protect\label{fig9}
\end{minipage}
\end{figure}

\newpage
{\large{\bf 2b. Charged Gauss-Bonnet Black Hole}}

The charged Gauss-Bonnet black hole has the following form of $f(r)$
\cite{wiltsh}: 
\be
f(r)=1+\frac{r^2}{2\alpha}-\frac{r^2}{2\alpha}\sqrt{1+\frac{8\alpha
    M}{r^{d-1}}-\frac{4\alpha Q^2}{2\pi (d-2)(d-3)r^{2d-4}}},
\ee
if $M>0$ and $\alpha >0$, then there will be a timelike singularity
which will be shielded by two horizons if $Q<Q_{ex}$. Here $Q_{ex}$ is
the extremal value of the charge determined from \cite{wiltsh}:
\be
r_{ex}^{2(d-3)}+\frac{d-5}{d-3}\alpha
r_{ex}^{2(d-4)}-\frac{Q_{ex}^2}{2\pi (d-2)(d-3)}=0,
\ee
where,
\be
r_{ex}^{d-3}=-\frac{1}{2}(d-5)M+\left[\frac{1}{4}(d-5)^2M^2+\frac{(d-4)Q_{ex}^2}{2\pi(d-2)(d-3)}\right]^{1/2}.
\ee
\begin{table}[here]
\begin{center}
\begin{tabular}[b]{|c|c|c|}  \hline \hline
$Q$ & \multicolumn{2}{|c|}{$\omega$}\\ \cline{2-3} & d=7 & d=8  \\
\hline
1.0  & 2.16438-0.44078i & 2.63435-0.54628i  \\
2.0  & 2.16556-0.43951i & 2.63485-0.54543i  \\
3.0  & 2.16750-0.43731i & 2.63566-0.54400i  \\
4.0  & 2.17016-0.43411i & 2.63677-0.54196i  \\
5.0  & 2.17347-0.42977i & 2.63814-0.53929i  \\
6.0  & 2.17730-0.42414i & 2.63971-0.53596i  \\
7.0  & 2.18137-0.41711i & 2.64143-0.53194i  \\
\hline\hline
\end{tabular}
\caption{Quasinormal frequency of charged Gauss Bonnet Black Hole for
$\alpha=1$, $l=2$ and $n=0$ in $d=7, 8$ for tensor perturbations.}
\end{center}
\end{table}

For $Q=Q_{ex}$, there is a single degenerate horizon at $r=r_{ex}$.
For the charged case, we find that the real part of the frequency is
just increasing and the damping is decreasing as the charge of the
black hole is increased. In Table 8, we give the quasinormal
frequencies of the charged Gauss-Bonnet black hole with the charge of
the black hole (normalized by the extremal value of the charge) for
$\alpha=1$, $l=2$ and $n=0$. 

In Figure (10) and (11) the real and imaginary part of the quasinormal
frequency is plotted as a function of $Q/Q_{ex}$. This nature is also
quite similar to that obtained by Konoplya \cite{konoplya} while
discussing the scalar perturbation of the charged Gauss-Bonnet black
hole.

\begin{figure}[!]
\begin{minipage}[t]{8cm}
\vspace{-10pt}
\centerline{\hspace{6.3mm}
\rotatebox{-90}{\epsfxsize=8cm\epsfbox{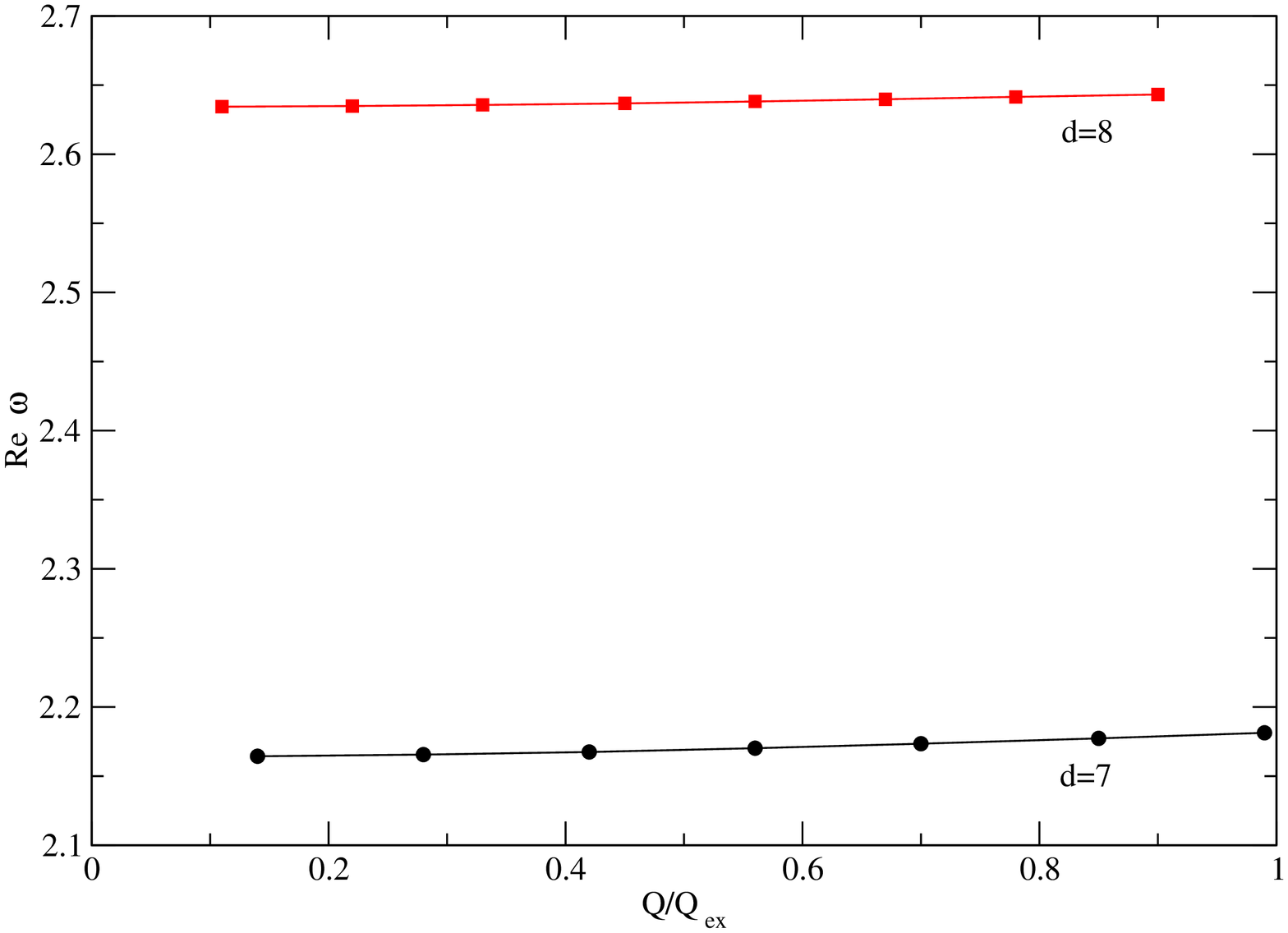}}}
\hspace{3.3cm}\caption[]{Plot of Re $\omega$ vs. $Q/Q_{ex}$ for
  charged Gauss-Bonnet black hole in $d=7$ and $8$ where $\alpha=1$, $l=2$ and $n=0$ }
\protect\label{fig10}
\end{minipage}
\hfill
\begin{minipage}[t]{8cm}
\vspace{-10pt}
\centerline{\hspace{11.3mm}
\rotatebox{-90}{\epsfxsize=8cm\epsfbox{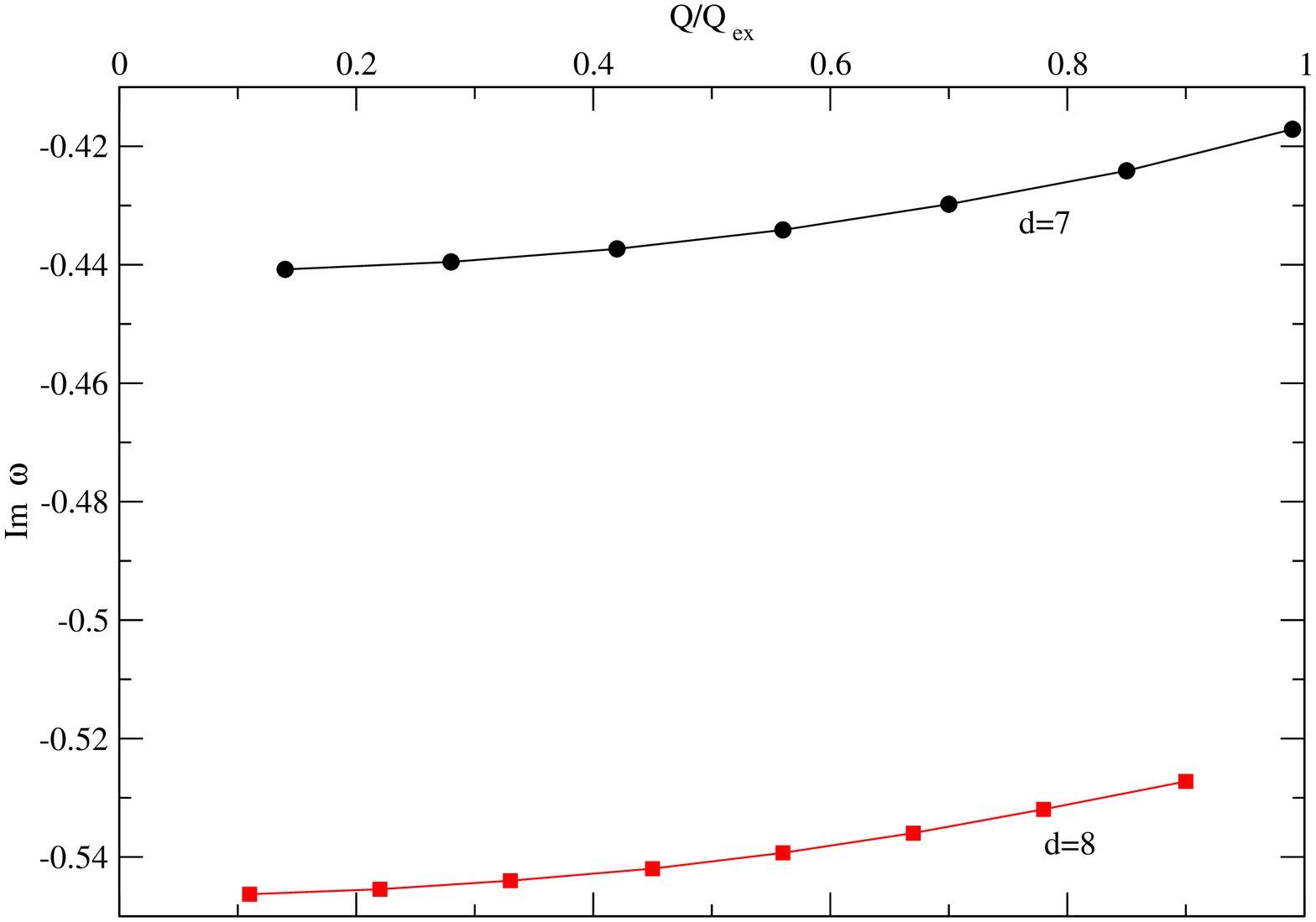}}}
\hspace{3.3cm}\caption[]{Plot of Im $\omega$ vs. $Q/Q_{ex}$ for
  charged Gauss-Bonnet black hole in $d=7$ and $8$ where $\alpha=1$, $l=2$ and $n=0$}
\protect\label{fig11}
\end{minipage}
\end{figure}

\sxn{Discussion}

We have obtained the QN frequencies for tensor type perturbations of
the uncharged Gauss-Bonnet black hole in five, seven and eight
dimensions using the tensor type gravitational potential derived in
\cite{dotti}. We have also obtained the QN frequencies for vector type
perturbations \cite{dottivector} of the same black hole in five, six,
seven and eight dimensions. We have shown that in the limit $\alpha\to
0$, one gets the QN frequency for the tensor type as well as vector
type perturbations of the Schwarzschild black hole. This indicates
that when the Gauss-Bonnet coupling $\alpha$ goes to zero, results of
Einstein gravity can be reproduced. The case for $d=6$ is ruled out in
case of tensor type perturbations because of the fact that it is
unstable to tensorial perturbations \cite{dotti}. We observe that the
real and imaginary part of the QN frequencies depend on the
Gauss-Bonnet coupling $\alpha$ for both the cases. The real
oscillation frequency for tensor and vector type perturbations always
increases with the increase of the GB coupling $\alpha$. The damping
decreases as we increase the Gauss-Bonnet coupling, but after reaching
certain value of $\alpha$ the damping increases. Though our results
for the QN frequencies of tensor perturbations matches qualitatively
with the results for quasinormal frequencies arising due to scalar
field perturbations \cite{konoplya}, the values of the QN frequencies
are different.

We have also discussed about the nature of QN frequency of the charged
GB black hole under tensor perturbations. The QN frequencies of the
charged Gauss-Bonnet black hole for vector perturbations were
difficult to find out because of the computational complicacy. It has
been found that the real part of the frequency increases and the
imaginary part decreases with the increase of charge for the tensor
perturbation case.

The problem of finding asymptotically highly damped QNMs is beyond
our study. The study of asymptotic QNMs of higher dimensional
Schwarzschild black hole with a Gauss-Bonnet correction was done in
\cite{ksg}, but the investigation of asymptotic QNMs for the full
Gauss-Bonnet metric still remains incomplete.

\sxn{Acknowledgement}

The author wishes to thank Kumar S. Gupta for numerous discussions
and comments during the work and for carefully reading the manuscript. 

\bibliographystyle{unsrt}

\begin{thebibliography}{99}


\bibitem{kk} Kokkotas, K.D., Schmidt, B.G.: {\it Living Rev. Rel.}
  {\bf 2}, 2 (1999) 
\bibitem{noll} Nollert, H.-P.: {\it Class. Quantum Grav.} {\bf 16},
  R159 (1999)
\bibitem{rg} Regge, T., Wheeler, J.A.: {\it Phys. Rev.} {\bf 108}, 1063
  (1957)
\bibitem{zerilli} Zerilli, F.J.: {\it Phys. Rev.} {\bf D2}, 2141 (1970)
\bibitem{vish} Vishveshwara, C.V.: {\it Phys. Rev.} {\bf D1}, 2870 (1970)
\bibitem{vish3} Vishveshwara, C.V.: {\it Nature}, {\bf 227} 936 (1970)
\bibitem{sterio} Kokkotas, K.D., Stergioulas, N.: Proc. $5$th
  Int. Workshop on New Worlds in Astroparticle Physics, Faro,
  Portugal, $8$-$10$ January, 2005, eds. Mourao, A.M. et al (World
  Scientific, Singapore, 2006) 
\bibitem{danny} Birmingham, D., Sachs, I., Solodukhin, S.N.:
{\it Phys. Rev. Lett.} {\bf 88}, 151301 (2002)
\bibitem{danny1} Birmingham, D., Sachs, I., Solodukhin, S.N.: {\it
    Phys. Rev.} {\bf D67}, 104026 (2003) 
\bibitem{hod} Hod, S.: {\it Phys. Rev. Lett.} {\bf 81}, 4293 (1998) 
\bibitem{dreyer} Dreyer, O.: {\it Phys. Rev. Lett.} {\bf 90}, 081301
  (2003)  
\bibitem{motl} Motl, L., Neitzke, A.: {\it Adv. Theor. Math. Phys.}
  {\bf 7}, 307 (2003) 
\bibitem{sdas} Das, S., Shankaranarayanan, S.: {\it Class. Quantum
  Grav.} {\bf 22}, L7 (2005)
\bibitem{sdas1} Ghosh, A., Shankaranarayanan, S., Das, S.: {\it
    Class. Quantum Grav. } {\bf 23}, 1851 (2006) 
\bibitem{natario} Nat\'{a}rio, J., Schiappa, R.: {\it
    Adv. Theor. Math. Phys.} {\bf 8}, 1001 (2004)
\bibitem{asen} Sen, A.: {\it JHEP} {\bf 0603}, 008 (2006) 
\bibitem{moura} Moura, F., Schiappa, R.: {\it Class. Quantum Grav. }
  {\bf 24}, 361 (2007)
\bibitem{scherk} Scherk, J., Schwarz, J.H.: {\it Nucl. Phys.} {\bf
    B81}, 118 (1974) 
\bibitem{zwei} Zwiebach, B.: {\it Phys. Lett} {\bf B156}, 315 (1985)
\bibitem{deser} Boulware, D.G., Deser, S.: {\it Phys. Rev. Lett.} {\bf
  55}, 2656 (1985) 
\bibitem{wheeler} Wheeler, J.T.: {\it Nucl. Phys.} {\bf B268}, 737 (1986)
\bibitem{wheeler1} Wheeler, J.T.: {\it Nucl. Phys.} {\bf B273}, 732 (1986)
\bibitem{wiltsh} Wiltshire, D.L.: {\it Phys. Rev.} {\bf D38}, 2445 (1988)
\bibitem{fromross} Meissner, K.A., Olechowski, M.: {\it Phys. Rev.}
  {\bf D65}, 064017 (2002)
\bibitem{ross} Cvetic, M., Nojiri, S., Odintsov, S.D.: {\it Nucl.
    Phys.} {\bf B628}, 295 (2002);
\bibitem{nojiri} Nojiri, S., Odintsov, S.D., Ogushi, S,: {\it
    Phys. Rev.} {\bf D65}, 023521 (2002)
\bibitem{cho} Cho, Y.M., Neupane, I.P.: {\it
    Phys. Rev.} {\bf D66}, 024044 (2002);
\bibitem{neu} Neupane, I.P.: {\it Phys. Rev.}
  {\bf D67}, 061501 (2003)
\bibitem{cai} Cai, R.G.: {\it Phys.Lett.} {\bf B582}, 237 (2004)
\bibitem{clunan} Clunan, T., Ross, S.F., Smith, D.J.: {\it Class.
    Quantum Grav.}  {\bf 21}, 3447 (2004)
\bibitem{alex} Barrau, A., Grain, J., Alexeyev, S.O.: {\it Phys.
    Lett.} {\bf B584}, 114 (2004)
\bibitem{vish1} Iyer, B.R., Iyer, S., Vishveshwara, C.V.: {\it Class.
  Quantum Grav.} {\bf 6}, 1627 (1989)
\bibitem{konoplya} Konoplya, R.: {\it Phys. Rev.} {\bf D71}, 024038 (2005)
\bibitem{kon1} Abdalla, E., Konoplya, R.A., Molina, C.: {\it Phys.
    Rev.} {\bf D72}, 084006 (2005)
\bibitem{iyer} Iyer, S.: {\it Phys. Rev.} {\bf D35}, 3632 (1987)
\bibitem{dotti} Dotti, G., Gleiser, R.J.: {\it Class. Quantum
    Grav.} {\bf 22} L1, (2005)
\bibitem{dottivector}  Gleiser, R.J., and Dotti, G.: {\it Phys. Rev.} {\bf
    D72}, 124002 (2005)
\bibitem{higuchi} Higuchi, A.: {\it J. Math. Phys.} {\bf 28}, 1553 (1987)
\bibitem{rubin} Rubin, M.A., Ord\'{o}\~{n}ez, C.R.: {\it J. Math.
    Phys.} {\bf 25}, 2888 (1984)
\bibitem{ishi} Ishibashi, A., Kodama, H.: {\it Prog. Theor. Phys.}
  {\bf 110}, 701 (2003)
\bibitem{ishi1} Kodama, H., Ishibashi, A.: {\it Prog. Theor. Phys.}
  {\bf 110}, 901 (2003)
\bibitem{chandra} Chandrasekhar, S., Detweiler, S.: {\it Proc. Roy.
    Soc.(London)} {\bf A344}, 441 (1975)
\bibitem{ferrari} Ferrari, V., Mashhoon, B.: {\it Phys. Rev.} {\bf
    D30}, 295 (1984)
\bibitem{will} Schutz, B., Will, C.M.: {\it Astrophys. J.} {\bf
    291}, L33 (1988)
\bibitem{will2} Iyer, S., Will, C.M.: {\it Phys. Rev} {\bf D35}, 3621 (1985)
\bibitem{kon2} Konoplya, R.A.: {\it Phys. Rev.} {\bf D68}, 024018 (2003)
\bibitem{Andersson} Andersson, N.: {\it Proc. R. Soc. (London)} {\bf
    A439}, 47 (1992); 
\bibitem{and1} Andersson, N., Linnaeus, S.: {\it Phys. Rev.} {\bf
    D46}, 4179 (1992)
\bibitem{leaver} Leaver, E.W.: {\it Proc. R. Soc. (London)} {\bf
    A402}, 285 (1985)
\bibitem{cardosoyousi} Cardoso, V., Lemos, J.P.S., Yoshida, S.:
  {\it Phys. Rev.} {\bf D69}, 044004 (2004)
\bibitem{ksg} Chakrabarti, S.K., Gupta, K.S.: {\it Int. J. Mod.
    Phys.} {\bf A21}, 3565 (2006)
\bibitem{konoplya3} Konoplya, R.A.: {\it Phys. Rev.} {\bf D68}, 124017 (2003) 
\end{thebibliography}

\end{document}